\documentclass[a4paper,fleqn,useAMS,usenatbib]{mnras}

\pdfoutput=1

\usepackage{mathptmx}

\usepackage[T1]{fontenc}
\usepackage{ae,aecompl}

\usepackage[dvips]{graphicx}
\usepackage{amsmath}
\usepackage{amsfonts}
\usepackage{amssymb}
\usepackage{comment}
\usepackage[dvipsnames]{xcolor}
\usepackage[%
        ]{hyperref}
        
\hypersetup{
	colorlinks=true,	
	urlcolor=MidnightBlue,		
	pdfpagelabels=true,
	hypertexnames=true,
	plainpages=false,
	naturalnames=true,
	pdftitle={The SNR of idealised radial velocity signals},    
	pdfauthor={Kipping and Wang},     
	linkcolor=WildStrawberry,          
	citecolor=ForestGreen,        
}

\newcommand{\pdf}{ \mathrm{Pr} }

\newcommand{\starry}{{\tt starry}}
\newcommand{\wwwcoolworlds}{\href{https://zenodo.org/records/10729373?token=eyJhbGciOiJIUzUxMiJ9.eyJpZCI6IjY0OGNjNTY5LWRlZGItNGY4NS04NDMzLWVkN2Y5ZDI3ZmZmYiIsImRhdGEiOnt9LCJyYW5kb20iOiIzMzMzNjUzMGYyZjZkZDExOTk5NGFkZDc2MDAxMjZiNSJ9.qllJwxmuDDU8JLi9NiAyO_KJ4B0G4mi56YHhQLq57DzOIRAJoLDac_FglsI64E1cyuC-f7KLe9KHP2HristCFA}{this URL}}

\title[The SNR of idealised radial velocity signals]{The SNR of idealised radial velocity signals}
\author[Kipping \& Wang]{David Kipping$^{1}$\thanks{E-mail:
\href{mailto:dkipping@astro.columbia.edu}{dkipping@astro.columbia.edu}} \& Xian-Yu Wang$^{2}$\\
$^{1}$Dept. of Astronomy, Columbia University, 550 W 120th Street, New York, NY 10027, USA\\
$^{2}$Department of Astronomy, Indiana University, 727 East 3rd Street, Bloomington, IN 47405-7105, USA}

\date{Accepted . Received ; in original form }

\pubyear{2024}

\begin{document}
\label{firstpage}
\pagerange{\pageref{firstpage}--\pageref{lastpage}}
\maketitle

\begin{abstract}
One of the most basic quantities relevant to planning observations and assessing detection bias is the signal-to-noise ratio (SNR). Remarkably, the SNR of an idealised radial velocity (RV) signal has not been previously derived beyond scaling behaviours and ignoring orbital eccentricity. In this work, we derive the RV SNR for three relevant cases to observers. First, we consider a single mass orbiting a star, revealing the expected result that $\mathrm{SNR}\propto K \sqrt{T}$, where $T$ is the observing window, but an additional dependency on eccentricity and argument of periastron. We show that the RV method is biased towards companions with their semi-major axes aligned to the observer, which is physically intuitive, but also less obviously that the marginalised bias to eccentricity is negligible until one reaches very high eccentricities. Second, we derive the SNR necessary to discriminate eccentric companions from 2:1 resonance circular orbits, although our result is only valid for eccentricities $e\lesssim0.3$. We find that the discriminatory SNR is $\tfrac{9}{8} e^2 (1-e^2)^{-1/2}$ times that of the eccentric planet solution’s SNR, and is thus typically an order-of-magnitude less. Finally, we have obtained a semi-empirical expression for the SNR of the idealised Rossiter-McLaughlin effect, revealing the bias with respect to spin-orbit alignment angle. Our formula is valid to within 10\% accuracy in 95.45\% of the training samples used (for $b\leq0.8$), but larger deviations occur when comparing to different RM models.
\end{abstract}


\section{Introduction}
\label{sec:intro}

The radial velocity (RV) method has provided one of our primary windows into the discovery and characterisation of the exoplanet population \citep{lovis:2010,wright:2018}. The statistical properties of this ensemble constrain the formation and evolutionary pathways of exoplanets and thus are of particular interest
(e.g. \citealt{cumming:2008,wittenmyer:2011,bonfils:2013,clanton:2014,bennett:2021,pinamonti:2022}). Of course, any effort to study the inherent properties of an observed population must carefully consider the effects of detection bias.

Concerning bias in RV exoplanet surveys, much attention has been previously given to orbital eccentricities \citep{ford:2005,ford:2006,zakamska:2011,hara:2019}. However, we note that the biases discussed therein are not an observational bias per sey, but rather biases in the reported values of eccentricities that stem from assumptions made during the inference procedure. For example, as a positive-definite quantity, it was quickly recognised that common inference techniques, such as Markov Chain Monte Carlo (MCMC) methods, cannot natively sample in eccentricity and expect to obtain unbiased results - especially for near-circular orbits \citep{ford:2005}. To illustrate this, consider that a \citet{metropolis:1953} walker, seeded from some trial position close to $e=0$, cannot make most negative eccentricity proposals due to the boundary condition of the problem. In contrast, positive jumps are unimpeded, thereby producing a positive bias in $e$. The boundary condition can be remedied to some degree by re-parameterisation to the Lagrangian elements $e\sin\omega$ and $e\cos\omega$ \citep{ford:2005}, although this now induces a prior in eccentricity of $\pdf(e) \propto e$. To impose say, a uniform prior in $e$, it would be necessary to correct for this behaviour, for example via importance sampling \citep{ford:2006}. More conveniently, \citet{anderson:2011} suggest simply using $\sqrt{e}\sin\omega$ and $\sqrt{e}\cos\omega$ as modified Lagrangian elements. However, a resolution to the entire sampling debacle is simply to use a different sampler. For example, nested sampling \citep{skilling:2004} does not suffer from the same problem and one can readily sample in $e$ directly \citep{feroz:2011}.

Even these steps do not fully dissolve the positive definite bias though. Even unbiased eccentricity posteriors still have the positive-definite quantity, which leads to bias in reported summary statistics \citep{zakamska:2011}. For example, consider an eccentricity posterior with a mode at zero eccentricity. Because eccentricity is positive definite, the common approach of quoting the median of the posterior samples as a representative summary statistic will be offset from zero. A modern solution is to simply avoid summary statistics altogether for demographics analyses. \citet{hogg:2010} show that such approaches are always biased and instead suggest the use hierarchical Bayesian models which avoid this entire problem. Finally, we note that non-Gaussian noise can also introduce eccentricity bias, something we do not consider here, but direct the reader to \citet{hara:2019}.

None of these are truly a detection bias though - they have no bearing as to whether a truly eccentric planet is more or less likely to be detected by a given RV survey. The probability of detecting a given signal - also known as the completeness rate - is generally assumed to depend exclusively upon its signal-to-noise ratio (SNR). In general, we expect completeness to asymptotically approach zero in the limiting case of SNR$\to0$ and vice verse approach unity in the case of SNR$\to\infty$. The shape of this completeness curve will depend upon the details of the survey, but is typically a S-shaped, quasi-logistic function characterisable via injection-recovery experiments (e.g. \citealt{christiansen:2016}).

Since completeness depends upon SNR, and since SNR depends upon system parameters (e.g. orbital period, stellar mass, etc) then completeness depends upon the system parameters too. In this way, surveys are generally more complete towards certain choices of system parameters than others - which is to say they have a detection bias towards certain types of planets. Any effort to infer the true population parameters must account for such bias, and thus have a function for completeness probability as a function of SNR, and of course SNR as a function of system parameters. To our knowledge, the full SNR function of a RV signal is absent in the literature, although it is often assumed (incorrectly) to be $K/\sigma_0$, where $K$ is the RV semi-amplitude and $\sigma_0$ is the time-integrated uncertainty. The closest example of an attempted solution appears to be in Equation~(14) of \citet{gaudi:2022}, where the eccentricity is formally absent but the authors notes a ``relatively weak dependence on eccentricity for $e\leq0.5$''.

To address this, we consider the idealised case of a densely and uniformly sampled RV curve over a time span that is long compared to the orbital period, $P$. Whilst this is obviously an idealised case, the advantage of this approach is that the result is completely general; there's no need to ingest the specific set of time samples into the calculation. In actual population inference work, we recommend calculating SNR rigorously using the time samples for each object due to subtle effects. For example, \citet{cumming:2004} show how for a fixed number of measurements per orbital period, detection efficiency steeply drops off at high eccentricities since sparse time sampling can miss the large but brief RV spike that occurs near periastron. Nevertheless, we argue that the idealised SNR is still valuable to calculate, since it is a general and analytic formalism that more clearly reveals basic trends which broadly sculpt the RV method and compare to the biases afflicting other detection techniques.

\section{A Single Planet}
\label{sec:single}

\subsection{SNR of a planet induced radial velocity curve}

We start by considering the stellar radial velocity, $V$, as a function of time, $t$, in response to an orbiting planet \citep{lovis:2010}:

\begin{align}
V(t) &= \gamma - K \big( e \cos\omega + \cos(f+\omega) \big),
\label{eqn:V}
\end{align}

where $K$ is the radial velocity semi-amplitude, $\gamma$ is a constant offset, $e$ is the planet's orbital eccentricity, $\omega$ is the planet's argument of periastron and $f$ is the planet's true anomaly.

The SNR of any signal is always a comparative act. In this case, we define the SNR as that obtained when comparing $V(t)$ to a simple flat-line model of $V(t) = \gamma$ i.e. $\lim_{K\to0} V(t)$. The SNR is sensitive to the time window under consideration. For the moment, we consider the SNR of a single cycle of $[0,P]$ (but will generalise this result later). We assume homoscedastic, white noise in what follows, where $\sigma_0$ defines the time-integrated noise over the same time interval as the time unit basis. Accordingly, our results ignore the effects of time-correlated stellar activity which have become an increasing concern, especially for small signals of order of meters-per-second \citep{rajpaul:2015,dumusque:2018,cameron:2021}.

As noted in the introduction, we follow \citet{gaudi:2022} in assuming here (and throughout) the idealized case of a densely and uniformly sampled RV curve over a time span that is long compared to the orbital period, $P$. To calculate SNR, we follow the analytic prescription presented in \citet{kipping:2023}, specifically their Equation~(15), which here becomes

\begin{align}
\mathrm{SNR} =& \frac{1}{\sigma_0} \sqrt{\int_{t=0}^{P} \big(V(t) - \gamma\big)^2\,\mathrm{d}t},\nonumber\\
\qquad=& \frac{K}{\sigma_0} \sqrt{ \int_{t=0}^{P} \big( e\cos\omega + \cos(f+\omega) \big)^2 \,\mathrm{d}t },
\end{align}

where it should be noted that $f$ is a function of time, $t$. To make progress, we need use a substitution and integrate with respect to $f$, rather than $t$. Making such a substitution, we can write

\begin{align}
\mathrm{SNR} &= \frac{K}{\sigma_0} \sqrt{ \int_{t=0}^{P} \big( e\cos\omega + \cos(f+\omega) \big)^2 \frac{\mathrm{d}t}{\mathrm{d}f}\,\mathrm{d}f },\nonumber\\
\qquad&= \frac{K}{\sigma_0} \sqrt{ \int_{t=0}^{P} \big( e\cos\omega + \cos(f+\omega) \big)^2 \Big(\frac{P}{2\pi}\Big) \frac{(1-e^2)^{3/2}}{(1+e\cos f)^2}\,\mathrm{d}f }.
\end{align}

The above has a closed-form solution, given by

\begin{align}
\mathrm{SNR} &= \Big(\frac{K/\sqrt{2}}{\sigma_0}\Big) \Upsilon(e,\omega) \sqrt{P},
\label{eqn:SNRP}
\end{align}

where we define

\begin{align}
\Upsilon(e,\omega) &= \sqrt{1-e^2} \sqrt{ 1 - \Big(\frac{1-\sqrt{1-e^2}}{e}\Big)^2\cos(2\omega) }.
\label{eqn:upsilon}
\end{align}

Recall that we chose our time interval to be $[0,P]$ and that Equation~(\ref{eqn:SNRP}) reveals a dependency of $\mathrm{SNR} \propto \sqrt{P}$ i.e. proportional to the square-root of the time interval. If the time interval, $T$, equals an integer number of periods, we thus expect

\begin{align}
\mathrm{SNR} &= \Big(\frac{K/\sqrt{2}}{\sigma_0}\Big) \Upsilon(e,\omega) \sqrt{T}.
\label{eqn:SNRgen}
\end{align}

Of course, in general the time sampling will not be precisely an integer number of periods, and the SNR will deviate depending on the precise phasing used. Nevertheless, the result provides a useful approximation for parameterising the SNR without concern for the details of the precise phasing of the observations with respect to the orbital position.

\subsection{Eccentricity bias}

The radial velocity semi-amplitude, $K$, also has a dependency on eccentricity, since

\begin{align}
K &= \Big(\frac{2\pi G}{P}\Big)^{1/3} \frac{M_P \sin i}{(M_{\star} + M_P)^{2/3}} \frac{1}{\sqrt{1-e^2}}.
\end{align}

\begin{figure*}
\begin{center}
\includegraphics[width=18.0cm,angle=0,clip=true]{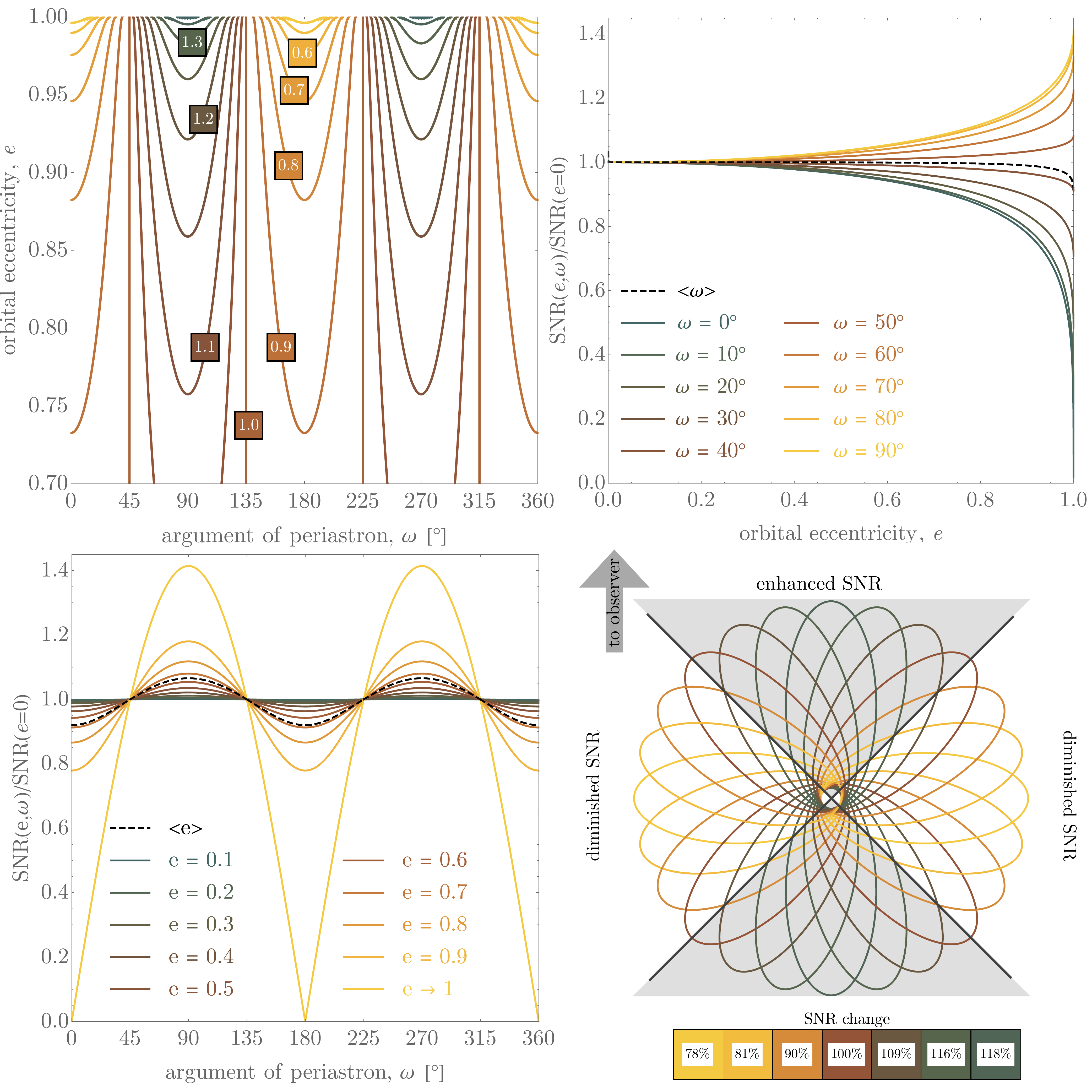}
\caption{\emph{
Corner plot of eccentricity bias.
Top-left: Contours of iso-SNR (holding everything else fixed) of a radial velocity planet, relative to the SNR of a circular orbit planet.
Top-right: Dependency of the bias (y-axis) upon eccentricity, for various choices of $\omega$. Note that because of the $\cos(2\omega)$ dependency, choices of $\omega>90^{\circ}$ wrap around.
Bottom-left: Same as top-right, but showing the dependency upon $\omega$ for various $e$ choices.
Bottom-right: Depiction of a $e=0.9$ orbit for various $\omega$ values, illustrating the effect on SNR.
}}
\label{fig:eccbias}
\end{center}
\end{figure*}

Accordingly, holding all else equal, we can define the bias with respect to $e$ and $\omega$ ($\mathcal{B}(e,\omega)$) as the fractional SNR change; i.e.

\begin{align}
\mathcal{B}(e,\omega) &= \frac{\mathrm{SNR}(e,\omega)}{\mathrm{SNR}(e=0)},\nonumber\\
\qquad&= \sqrt{ 1 - \Big(\frac{1-\sqrt{1-e^2}}{e}\Big)^2\cos(2\omega) },
\label{eqn:Bew}
\end{align}

where we note that $\lim_{e\to0} \mathcal{B}(e,\omega) = 1$, as expected. The behaviour of $\mathcal{B}(e,\omega)$ is illustrated in Figure~\ref{fig:eccbias}. Of particular interest is the limiting case of $\lim_{e\to1} \mathcal{B}(e,\omega) = \sqrt{1-\cos(2\omega)}$. In this limit, then, the SNR can fall to zero for planets with $\omega=0$ or $\pi$ - which corresponds to a 1D straight line ``orbit'' that's normal to the observer's line of sight and thus imparts zero radial velocity. It also reveals the maximum positive bias of $\sqrt{2}$, thereby bounding $0<\mathcal{B}(e,\omega)<\sqrt{2}$ for all $0\leq e < 1$.

We can now marginalize out the $\omega$ term under the assumption that $\omega$ is isotropically distributed in the cosmos, to give

\begin{align}
\mathcal{B}(e) =& \frac{\int_{\omega=0}^{2\pi} \mathcal{B}(e,\omega)\,\mathrm{d}\omega}{\int_{\omega=0}^{2\pi} \mathrm{d}\omega},\nonumber\\
\qquad=& \frac{\sqrt{2}}{\pi e} \Big(
\sqrt{1-(1-e^2)^{1/2}} E[1-\sqrt{1-e^2}] \nonumber\\
\qquad& + (1-e^2)^{1/4} \sqrt{1-\sqrt{1-e^2}} E[1 - (1-e^2)^{-1/2}]
\Big).
\label{eqn:Be}
\end{align}

Equation~(\ref{eqn:Be}) is a remarkably flat function with respect to eccentricity (see Figure~\ref{fig:eccbias}), revealing negligible bias with respect to eccentricity. What little dependency there is only really turns on at very high eccentricities. For example, the SNR is only decreased by $0.1$\% when the eccentricity reaches $e=0.63$, 1\% by $e=0.90$ and $5$\% by $e=0.994$. Taking the limit of extreme eccentricity, we find

\begin{align}
\lim_{e\to0} \mathcal{B}(e) &= 1,\nonumber\\
\lim_{e\to1} \mathcal{B}(e) &= \frac{2\sqrt{2}}{\pi} = 0.900...
\end{align}

\subsection{Argument of periastron bias}

Without marginalising out $\omega$, we can see the dependency of $\mathrm{SNR}$ upon $\omega$, or really $\cos(2\omega)$ (see Equation~\ref{eqn:Bew}). The $\cos(2\omega)$ is immediately interesting and can be understood by consideration of the orbital geometry. If $\omega=\pi/2$, the position of periastron is closest to the observer, whereas at $\omega=3\pi/2$ it is the position of apoastron. However, in either case, the semi-major axis lies along the observer's line of sight. The planet and star experience more motion along this axis than the semi-minor axis (simply by virtue of its greater length) and thus the radial velocity variations will be maximised along this direction. Accordingly, we should expect, from physical arguments, a bias towards $\omega=\pi/2$ and $3\pi/2$ for which we note $\cos(2\omega) = -1$ in both cases. In contrast, when $\omega=0$ or $\pi$, the semi-major axis is normal to the observer's line of sight, minimising the radial velocity variations. Again, those positions are described by a singular value of $\cos(2\omega)$ given by $\cos(2\omega)=+1$.

Another way to view the $\omega$ dependency is to marginalise over $e$ in Equation~(\ref{eqn:Bew}). Such an act implicitly assumes a uniform prior in eccentricity, which we know does not represent any physical population of planets \citep{beta:2013}. Nevertheless, this allows us to see the dependency on $\omega$ more explicitly, yielding

\begin{align}
\mathcal{B}(\omega) =& \frac{\int_{e=0}^{1} \mathcal{B}(e,\omega)\,\mathrm{d}e}{\int_{e=0}^{1} \mathrm{d}e},\nonumber\\
\qquad=& \sqrt{1-\cos2\omega}+2\sqrt{\cos2\omega}\sin^{-1}\sqrt{\cos2\omega} \nonumber\\
\qquad& - \frac{2\cos2\omega}{\sqrt{1+\cos2\omega}} \sin^{-1}|\cos\omega|.
\label{eqn:Bw}
\end{align}

As before, this behaviour is illustrated in Figure~\ref{fig:eccbias}, where we see the expected preference towards $\omega=\pi/2$ and $3\pi/2$ geometries. To our knowledge, this preference (whilst certainly intuitive) has not been rigorously demonstrated in the literature before. It is particularly interesting that this favours both the best and worst geometries for a transit to occur. Computing the resulting transit probabilities cannot be done without assuming some completeness curve, and thus we don't attempt to do so here, but highlight this curious feature.

Our expressions reveal the $e$-$\omega$ bias of idealised RV observations, which is particularly important for interpretting demographic trends. For example, should a survey report a pile-up of eccentric planets near $\omega=\pi/2$ and $3\pi/2$, this can be understood as purely a product of observational bias rather than an astrophysical phenomenon.

\section{Distinguishing Resonances}
\label{sec:double}

Another useful application of the SNR formalism is to the curious degeneracy highlighted in \citet{anglada:2010}. In that work, it is shown that the radial velocity signal caused by a single planet on an eccentric orbit can be well-fit by two planets in a 2:1 resonance. This is most easily seen by performing a series expansion of Equation~(\ref{eqn:V}) in low eccentricity, yielding

\begin{align}
V(t) &= \gamma - K \sin\Big(\frac{2\pi}{P} (t-\tau)\Big) - K e \sin\Big(\frac{2\pi}{(P/2)} (t-\tau) - \omega \Big) + \mathcal{O}[K e^2]
\end{align}

whereas two circular-orbit planets in a 2:1 resonance have

\begin{align}
V(t) &= \gamma - K_1 \sin\Big(\frac{2\pi}{P_1} (t-\tau_1)\Big) - K_2 \sin\Big(\frac{2\pi}{P_2} (t-\tau_2)\Big).
\end{align}

To first order in eccentricity, the degeneracy is exact. The only way to break this degeneracy with RVs alone is to seek evidence for the $\mathcal{O}[K e^2]$ term, which should not exist in the 2:1 resonance scenario. It is therefore necessary to include this term explicitly, which was solved for already in \citet{lucy:2005}:

\begin{align}
V(t) =& \gamma - K \sin\Big(\frac{2\pi}{P} (t-\tau)\Big) - K e \sin\Big(\frac{2\pi}{(P/2)} (t-\tau) - \omega \Big) \nonumber\\
\qquad& - \tfrac{9}{8} K e^2 \sin\Big(\frac{2\pi}{(P/3)} (t-\tau) - 2\omega \Big) + \mathcal{O}[K e^3].
\label{eqn:lucy}
\end{align}

The relevant SNR comparison now is the 2nd-order expansion versus the 1st-order expansion, such that 

\begin{align}
\mathrm{SNR}_{\mathrm{true\,\,ecc}} &= \frac{1}{\sigma_0} \int_{t=\tau}^{\tau+P} \Big( \frac{9}{8} K e^2 \cos\big(3\frac{2\pi (t-\tau)}{P}-2\omega\big) \Big)^2\,\mathrm{d}t.
\end{align}

Solving for the above and again making the generalisation we made before to arbitrary numbers of orbital periods, this yields

\begin{align}
\mathrm{SNR}_{\mathrm{true\,\,ecc}} &= \frac{9}{8} \Big(\frac{(K e^2)/\sqrt{2}}{\sigma_0}\Big) \sqrt{T}.
\label{eqn:SNRtru}
\end{align}

The above is conditional upon the second-order expansion in Equation~(\ref{eqn:lucy}). The maximum eccentricity to which this approximation is useful can be estimated considering the ratio of these first two terms ($\sum_{n=1}^{2} e^n$) against the sum of all terms ($\sum_{n=1}^{\infty} e^n$) and setting to some critical threshold which we choose to be 0.9:

\begin{align}
\frac{ \sum_{n=1}^{2} e^n}{ \sum_{n=1}^{\infty} e^n } &= 0.9.
\end{align} 

Solving the above for $e$ yields $e=0.316...$ and thus we suggest that the utility of Equation~(\ref{eqn:SNRtru}) is restricted to the range of $0 \leq e < 0.316$, since the expansion contains $>90\%$ of the power in this range.

\begin{figure}
\begin{center}
\includegraphics[width=8.4cm,angle=0,clip=true]{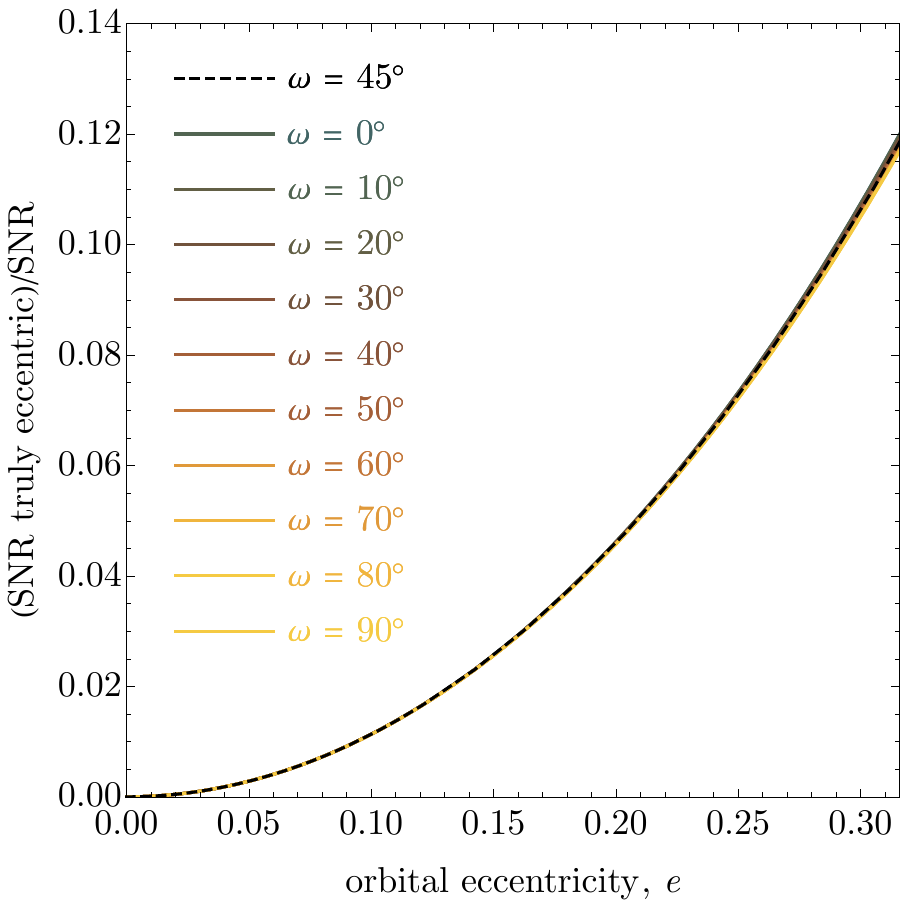}
\caption{\emph{
To discriminate between a 2:1 resonance and a truly eccentric planet, one needs to measure terms in the radial velocity signal of order $K e^2$. We plot this SNR, relative to the signal's total SNR, as a function of eccentricity.
}}
\label{fig:discrimination}
\end{center}
\end{figure}

A useful comparison at this point is to normalise this by the SNR of detecting an eccentric planet, in general i.e. Equation~(\ref{eqn:SNRgen}). This leaves us with

\begin{align}
\frac{ \mathrm{SNR}_{\mathrm{true\,\,ecc}} }{ \mathrm{SNR} }&= \frac{9}{8} \frac{e^3}{ \sqrt{(1-e^2)\big(e^2-(1-\sqrt{1-e^2})^2\cos(2\omega)\big)} },
\end{align}

which we plot in Figure~\ref{fig:discrimination}. As before, $\omega$ only appears within the $\cos(2\omega)$ container and thus we only need plot $\omega=[0,\pi/2]$ to see the full range of possible behaviours. In practice, the $\omega$ dependency is extremely weak for $e<0.316$ and thus one may simply use

\begin{align}
\frac{ \mathrm{SNR}_{\mathrm{true\,\,ecc}} }{ \mathrm{SNR} }&\simeq \frac{9}{8} \frac{e^2}{\sqrt{1-e^2}}.
\label{eqn:anglada}
\end{align}

We suggest interpreting the above and Figure~\ref{fig:discrimination} as follows. If one has an RV signal that appears eccentric with some SNR value, the expression can be used to read off the expected SNR of the resonance-vs-eccentric discrimination term, relative to the full SNR. For example, if $e=0.2$, then the discrimination term will have an SNR 4.6\% that of the full SNR. So if the full SNR is, say 25, the discrimination term will be at $\mathrm{SNR}_{\mathrm{true\,\,ecc}} = 1.1$ and is thus likely inadequate to make this discrimination with much confidence.

At first blush, it may appear strange that the discrimination power tends to zero for a circular orbit, as shown in Figure~\ref{fig:discrimination}. The explanation is in fact rather simple. For $e=0$, the Fourier expansion of the RV curve has no power for the $K e$ terms, which equal $K_2$ in the 2:1 degeneracy scenario. In other words, a single planet with period $P$ and radial velocity amplitude $K$ is perfectly described by a 2:1 resonance with the first planet at $P_1=P$ and $K_1=K$ and the second planet at $P_2=P/2$ and $K_2=0$. This is of course obvious in hindsight, for any circular orbit planet, one cannot rule out the possibility that there is a second planet with negligible mass - that's always a possibility.

\section{Rossiter-McLaughlin Effect}
\label{sec:RM}

\subsection{The need for a numerical approach}

In Section~\ref{sec:single}, the SNR could be expressed exactly and in closed-form. In Section~\ref{sec:double}, we again found a closed-form solution, but invoked a Fourier series expansion to make progress. Finally, in this section, we will use a semi-empirical approach guided by numerical simulations to explore the SNR of the Rossiter-McLaughlin (RM) effect.

To calculate a closed-form SNR, the integral of the underlying model squared must yield an analytic solution. For the RM effect, even writing down a closed-form solution for the integrand is challenging, with many forward model packages electing to use a numerical approach instead (e.g. \starry; \citealt{lugur:2019}). Analytic models, under various approximate conditions, do exist \citep{ohta:2005,hirano:2011,sasaki:2021}. However, we were unable to obtain useful analytic expressions with any of these models and thus turn to numerical simulations across a grid of possible inputs. These results will then we used to deduce semi-empirical scaling relations.

\subsection{Known scaling behaviours}

Before we detail our numerical simulations, it is worth first discussing what is already known about the scaling behaviour of the RM effect. From \citet{gaudi:2007}, one can show that the RM amplitude, $K_{RM}$, scales as

\begin{align}
K_{RM} = V \sin I_{\star} \frac{p^2}{1-p^2},
\label{eqn:KRM}
\end{align}

where $p$ is the ratio of the planet to star radius, $V$ is the equatorial rotation speed of the stellar photosphere abd $I_{\star}$ is the inclination of the stellar spin axis relative to the sky plane. This insight immediately tells us that our numerical grid need not sample different values of $V \sin I_{\star}$ nor $p$ - the dependency is already established.

What parameters should be included in the grid then? We will certainly need to vary the sky-projected spin-orbit angle, $\lambda$, in our grid - which critically affects the shape of the RM curve. Guided by consideration of the analogous transit problem \citep{kipping:2016}, we expect the first-to-fourth and second-to-third contact transit durations ($T_{14}$ \& $T_{23}$) to have a major effect on the SNR. In particular, one might expect some kind of duration to the index of one-half scaling, for idealised noise statistics.

Many other parameters could also be included at this stage, such as limb darkening, micro-turbulence, macro-turbulence, etc. However, these are all second-order effects and our primary interest here to derive a simple and approximate SNR equation to maximise its accessibility and utility. With this limitations in mind, we proceed with a three-dimensional grid as described above.

\subsection{Generating a grid of RM simulations}
\label{sub:RMsims}

The grid is chosen to span $\lambda = [0^{\circ},350^{\circ}]$ in $10^{\circ}$ steps, impact parameters of $b = [0,1]$ in $0.1$ steps and $\log_2 (a/R_{\star}) = [1,8]$ in 0.25 steps. Thus, a total of 11,484 unique grid positions were defined. In each case, $(a/R_{\star})$ and $b$ are converted to $T_{14}$ and $T_{23}$ using the \citet{seager:2003} equations, assuming a circular orbit and a Solar bulk stellar density to obtain orbital period, $P$.

We present two grids of RM simulations based on the widely utilized models by \citet{ohta:2005} and \citet{hirano:2011}. The model by \citet{ohta:2005} calculates the intensity-weighted mean wavelength of distorted spectral lines, offering a straightforward approach but lacking in high accuracy. Conversely, the model by \citet{hirano:2011} employs a more sophisticated method that incorporates realistic effects, including rotational broadening, macroturbulence, and other subtle phenomena.

The models by \citet{ohta:2005} and \citet{hirano:2011} have been implemented in \texttt{ellc} \citep{Maxted:2016} and \texttt{tracit} \citep{Knudstrup:2022, Hjorth:2021}, respectively, which are adopted as RM generators. In the grid generation process, we set the projected rotational velocity ($v \sin i$), radius ratio ($R_p/R_*$), eccentricity ($e$), and the argument of periastron ($\omega$) to 3 km/s, 0.1, 0, and 0, respectively. Furthermore, to exclude second-order effects, we fixed the limb-darkening coefficients, micro-turbulent, and macro-turbulent velocities to zero.

\subsection{Linearised RM effect}

To make progress, we decided to approximate the RM curves as a polygon pulse, following analogous studies for transits \citep{carter:2008}. The pulse has a value of $\Delta V_{RM}(t)=0$ for all $t<-T_{14}/2$ and all $t>T_{14}/2$. During the ingress, the RM model is a linear line apexing at $\Delta V_{RM}(-T_{23}/2)=V_1$. It then takes on a new trend such that it reaches the second apex at $\Delta V_{RM}(T_{23}/2)=V_2$. The egress then returns to zero at the fourth contact. This linearised approximation thus takes the form

\begin{equation}
\Delta V_{RM}(t) \simeq
\begin{cases}
0 & \text{if } t \leq -T_{14}/2,\\
\frac{2 V_1 (t+(T_{14}/2)}{T_{14}-T_{23}} & \text{if } -T_{14}/2 < t \leq -T_{23}/2,\\
V_1 - \frac{V_1-V_2}{2} - \frac{(V_1-V_2)t}{T_{23}} & \text{if } -T_{23}/2 < t \leq T_{23}/2,\\
-\frac{2V_1 (t-(T_{14}/2))}{T_{14}-T_{23}} & \text{if } T_{23}/2 < t \leq T_{14}/2,\\
0 & \text{if } t > T_{14}/2,
\end{cases}
\label{eqn:linearRM}
\end{equation}

\begin{figure*}
\begin{center}
\includegraphics[width=18.0cm,angle=0,clip=true]{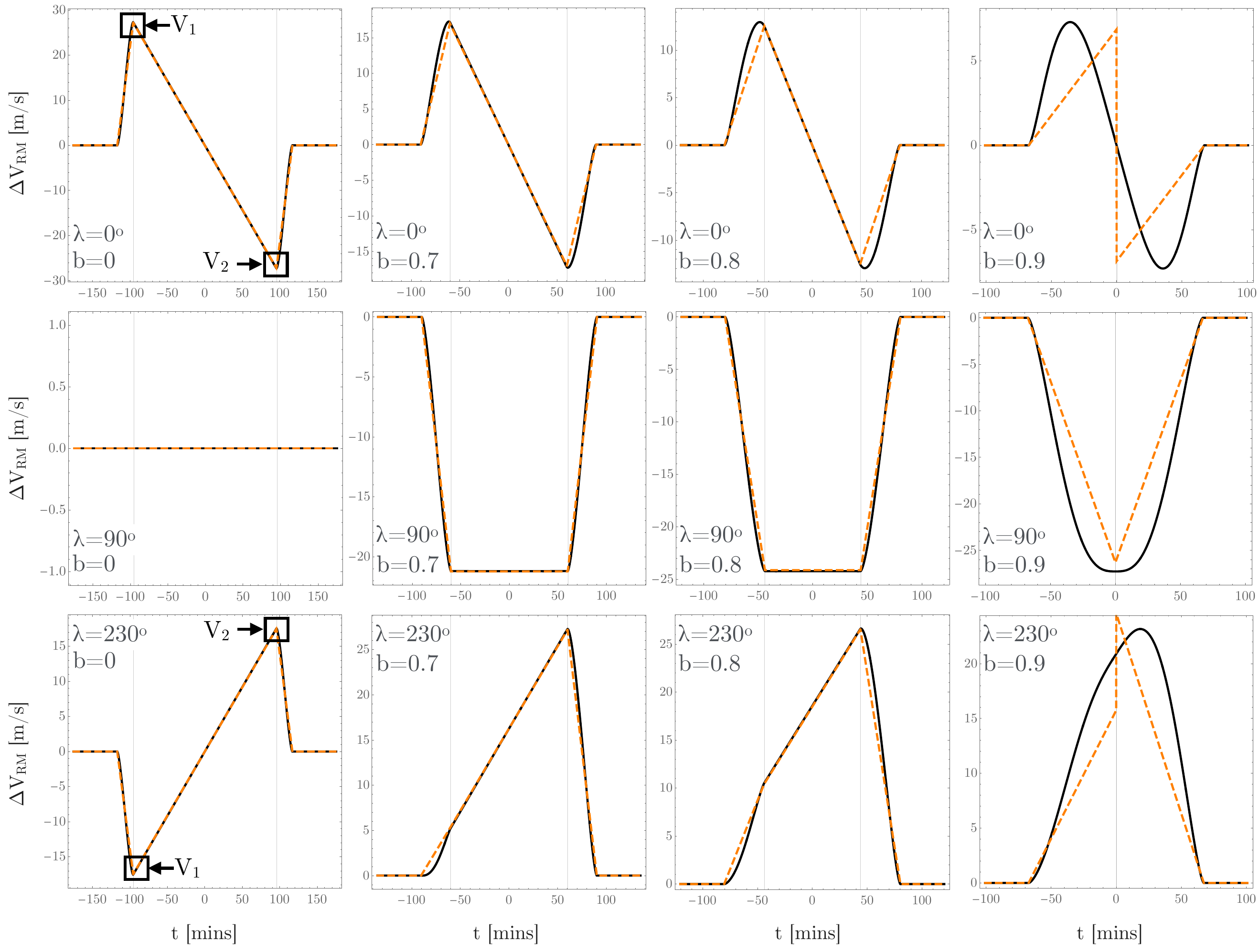}
\caption{\emph{
Grid of RM curves simulated as described in Section~\ref{sub:RMsims} (black solid lines). In all simulations $p=0.1$ and $a_R=16$, although we find only the scaling is affected by these parameters. The RM curves are approximated by a linearised function (orange-dashed) with discontinuities at $t=-T_{23}/2$ and $t=T_{23}/2$, for which the velocities are $V_1$ and $V_2$, respectively. This approximation breaks down for $b>0.8$.
}}
\label{fig:linearRM}
\end{center}
\end{figure*}

This linearized RM curve well approximates the typical RM examples for $b \leq 0.8$, whereas for high impact parameters, the linear model breaks down (See Figure~\ref{fig:linearRM}). We were unable to find a simple extension for these cases and thus simply discard them in what follows. Therefore, our eventual semi-empirical will only be valid for $b \leq 0.8$.

\subsection{Analytic SNR}

\begin{figure}
\begin{center}
\includegraphics[width=8.4cm,angle=0,clip=true]{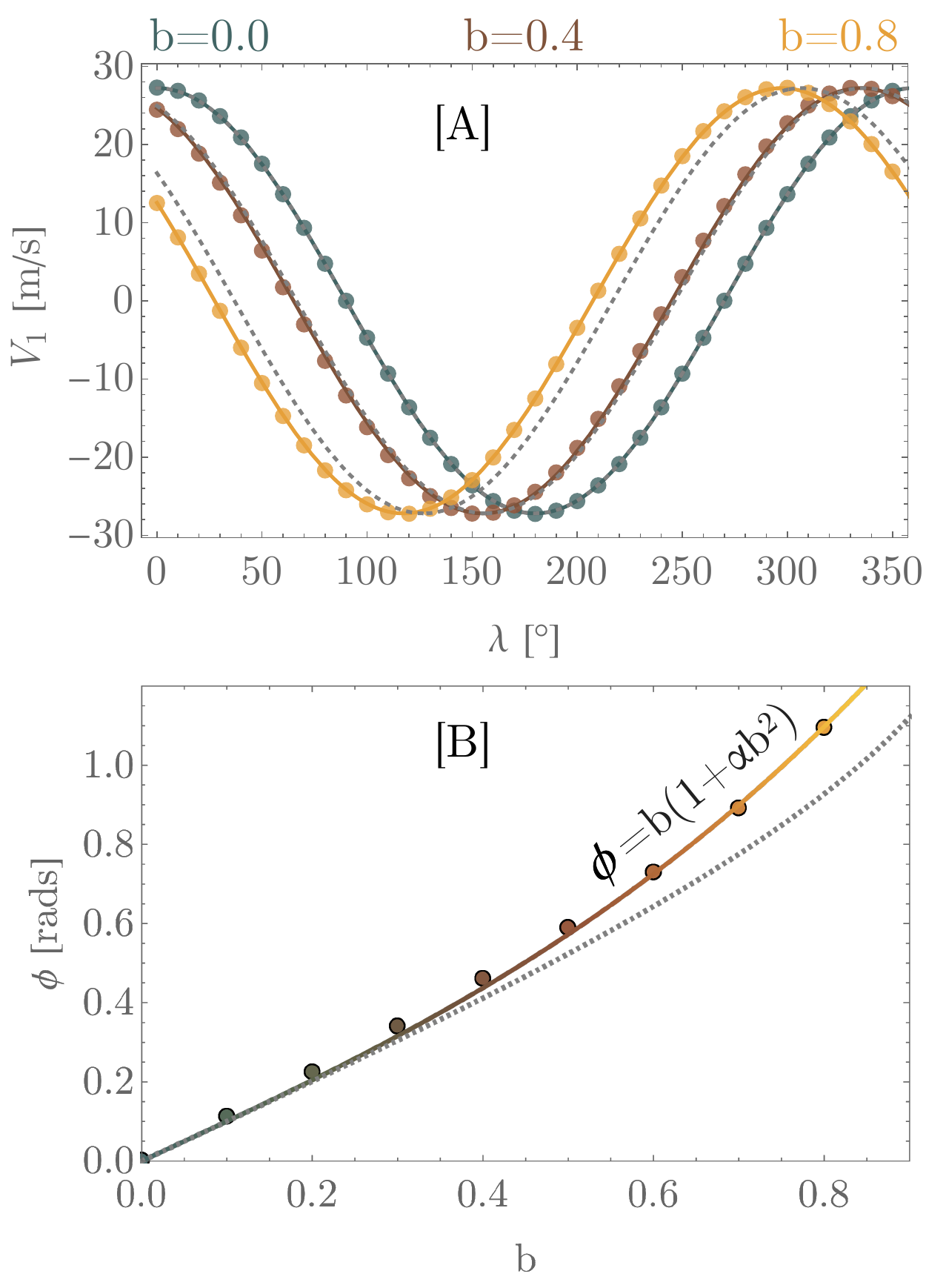}
\caption{\emph{
A] Dependency of $V_1$ versus $\lambda$ for three choices of the impact parameter, $b$ (circles). The dependency appears well-described by a cosine function (solid line), with an increasing phase shift, $\phi$, as $b$ grows. The gray dashed lines show an alternative expression for $V_1$ presented in \citet{albrecht:2022}.
B] Points shows the best-fitting phases for a grid of impact parameters, $b$. The annotated function appears to well-describe the relationship. Again, the dotted line shows the phase predicted by the alternative formula of \citet{albrecht:2022}.
}}
\label{fig:V1}
\end{center}
\end{figure}

Our next task is to find expressions for $V_1$ and $V_2$ as a function of the input parameters. Plotting $V_1$ as a function of $\lambda$ and fixed $a_R$ and $b$, we first observe that this relation is independent of $a_R$. This can be understood by the fact that $a_R$ primarily governs the speed of the planet across the stellar disk, and thus does not affect the amplitude of the RM effect itself. For $b=0$, $V_1 \propto \cos\lambda$ provides an ostensibly perfect description of the simulation results. We observe that increasing $b$ shifts the cosine over i.e. a phase term, $\phi$, is added. This is shown in Figure~\ref{fig:V1}A.

For each $b$ choice, we fit for the phase term in a least squares sense and then plotted the resulting phases versus impact parameter, as seen in Figure~\ref{fig:V1}B. We find that the function $\phi = b(1+0.58b^2)$ provides an excellent match to the observed behaviour and thus we determine that

\begin{align}
V_1 &= K_{RM} \cos\big(\lambda+b(1+\alpha b^2)\big),
\end{align}

where $\alpha=0.58$ and $K_{RM}$ is the amplitude (given earlier in Equation~\ref{eqn:KRM}), which we'll return to shortly. We repeated these experiments for $V_2$ and find very similar results except with sign changes such that

\begin{align}
V_2 &= -K_{RM} \cos\big(\lambda-b(1+\alpha b^2)\big).
\end{align}

We note that \citet{albrecht:2022} previously presented alternative ecpressions for $V_1$ and $V_2$ in their Section~3.1.1, which are

\begin{align}
V_2 + V_1 &= 2 V \sin I_{\star} \cos\lambda \sqrt{1-b^2},\\
V_2 - V_1 &= 2 V \sin I_{\star} \sin\lambda b.
\end{align}

These expressions are equivalent to our own in the limit of $b\to0$ and also enjoy a more physically motivated justification than our empirical formulae. However, we find that they do not provide a good match to grid of simulations as $b$ grows, which can be seen in Figure~\ref{fig:V1}. Accordingly, we decided to use our own $V_1$ and $V_2$ expressions in what follows.

Equipped with these expressions, we can now integrate the square of Equation~(\ref{eqn:linearRM}) to obtain the final SNR expression, given by

\begin{align}
\mathrm{SNR}_{RM} =& \frac{K_{RM}}{\sqrt{6}\sigma_0} \Bigg( T_{14}+T_{23}\big(1-\cos[2\lambda]\big)+\cos[2(b+\alpha b^3)] \nonumber\\
\qquad& \times \Big( (T_{14}+T_{23})\cos[2\lambda] - T_{23} \Big) \Bigg)^{1/2}.
\label{eqn:SNRRM}
\end{align}

\begin{figure}
\begin{center}
\includegraphics[width=8.4cm,angle=0,clip=true]{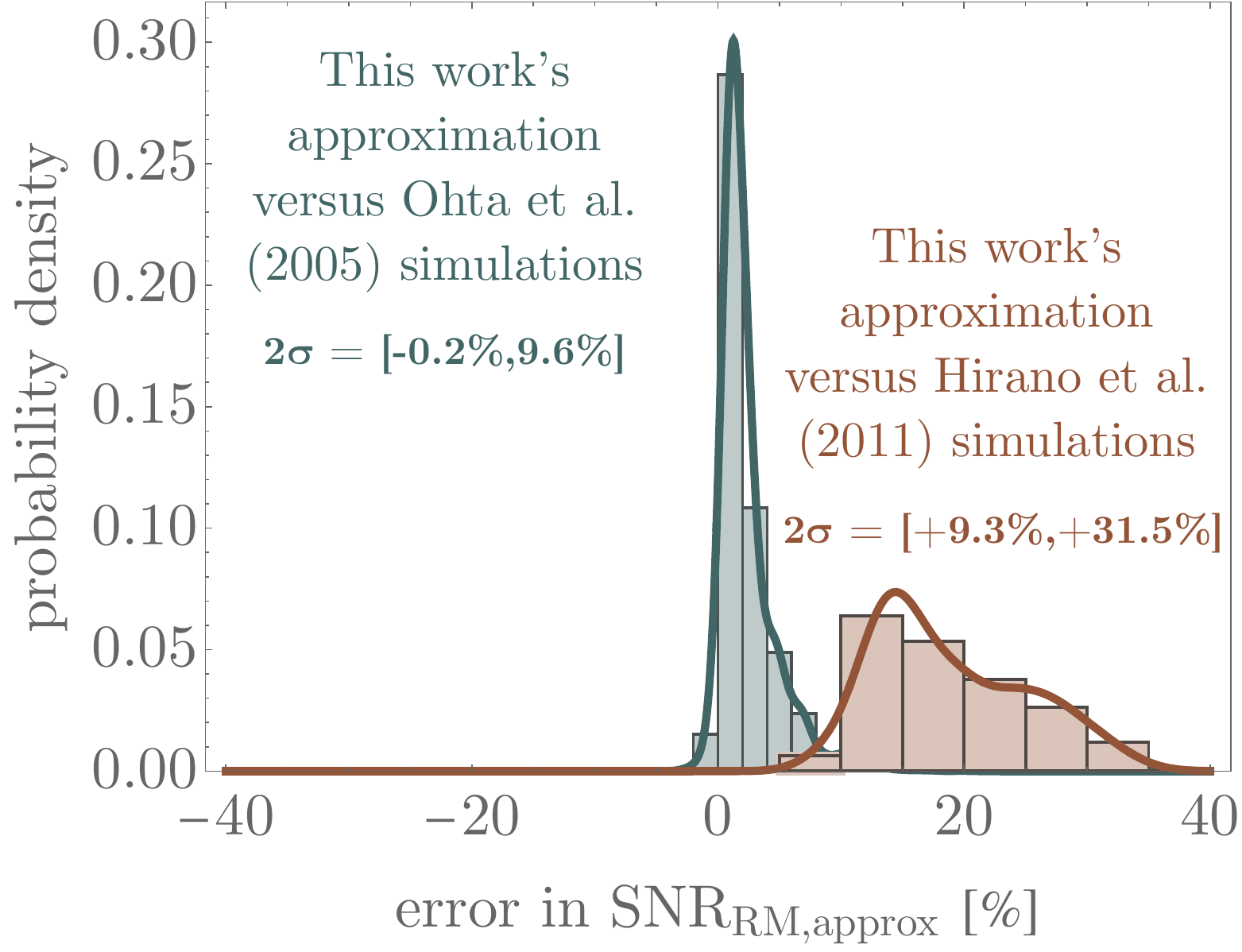}
\caption{\emph{
Comparison of the SNR for the RM effect calculated using our approximate formula of Equation~(\ref{eqn:SNRRM}) versus the exact numerical solution using the \citet{ohta:2005} simulations used throughout (green). For 95.45\% of experiments, the error is within 10\%, although we note a tendency for our formula to systematically overestimate the SNR. For comparison, we show the same formula applied to identical set of simulations but using the \citet{hirano:2011} model, where we note these simulations produced 20\% systematically higher SNR values than predicted using Equation~(\ref{eqn:SNRRM}).
}}
\label{fig:SNRacc}
\end{center}
\end{figure}

We compared the resulting SNRs using Equation~(\ref{eqn:SNRRM}) to the exact numerical answer found from the simulated RM curves. This was achieved by performing a spline interpolation of the curves and then integrating the square of these functions over time. We plot the relative deviance of the approximate formula to the true values in Figure~\ref{fig:SNRacc}. From our 9338 grid simulations with $b\leq0.8$ and $\mathrm{SNR}_{RM}>0$, we find the worst case is a 20\% offset between the two. In general, the median of the comparisons is a 1.6\% error with Figure~\ref{fig:SNRacc} revealing a general tendency of our formula to slightly overestimate the true SNR. However, 95.45\% of the experiments are within 10\% error and thus we consider the formula a useful tool for the community, given the apparent total absence of any alternative at this point.

We envisage our equations as being useful in population-level analyses of RM observations. For example, one might find that more RM detections are obtained in a certain part of parameter space than elsewhere and here our equations could reveal whether the excess was consistent with observational bias or not.

\section{Discussion}
\label{sec:discussion}

Three new results are presented in this work: 1) the RV SNR of a single companion, 2) the SNR necessary to discriminate the 2:1 degeneracy noted by \citet{anglada:2010}, and 3) an approximate SNR formula for the the RM effect. In all three cases, the SNR is calculated using the continuous integration technique described in \citet{kipping:2023}. Two important subtleties should be emphasised with this approach.

First, we explicitly assume idealised signals, in particular that the measurement uncertainties are normally and independently distributed. Such an assumption is appropriate when spurious signals from stellar activity or time-correlated instrumental noise are substantially less than the photon noise budget. However, many modern spectrographs are pushing down to sub-meters/second level where these time-correlated signals become dominant. Our approach is not valid for such cases, but nevertheless provides a starting point for future work to compare to.

Second, measurement noise appears in our formulae only through the term $\sigma_0$, which is not simply the typical uncertainty in a RV time series. Rather, it is the time-integrated error in the same time units as that used elsewhere in the formula. For example, suppose that one chooses to define $T_{23}$ and $T_{14}$ in hours. In this case, $\sigma_0$ would represent the time-integrated noise over one hour. So, if each measurement error was $\sigma$ with an exposure time of say 15\,minutes, then $\sigma_0 = \sigma/\sqrt{4}$. It is crucial $\sigma_0$ is treated properly in the formula to arrive at reliable SNR values.

Our work reveals some intuitive and less intuitive results. For the RV signal of a single companion, we find that orbits aligned with the semi-major axis pointed towards the observer enjoy a positive detection bias, which is intuitionally understood by maximising the radial component of the orbital motion. In contrast, a perhaps less intuitive result is that the SNR of a single companion, marginalised over all orientations, is almost flat. The RV method experiences very little change in sensitivity to even highly eccentric planets - averaged over all orientations. This useful result implies that the observed exoplanet eccentricity distribution from radial velocities is approximately unbiased, as is.

For the \citet{anglada:2010} resonance degeneracy, we find that the discrimination SNR is typically an order-of-magnitude below the observed signal's SNR. Hence, precise observations will be required to break this degeneracy using RVs alone. Our resulting expression (Equation~\ref{eqn:anglada}) provides the community with a quick and easy tool to estimate the required SNR, but we caution it's accuracy is limited to $e\lesssim0.3$.

The RM effect is the most challenging case to estimate the SNR for, since the models are simply too complex to be tractable with the integration method of \citet{kipping:2023}. We used the \citet{ohta:2005} model to simulate RM curves and then linearised them as an approximation, following \citet{carter:2008}. This results in an approximate and semi-empirical SNR formula in Equation~(\ref{eqn:SNRRM}). Numerical experiments show the formula is accurate to within 10\% in 95.45\% of cases. However, that's only a comparison of the formula to its own training data - the \citet{ohta:2005} model. In Figure~\ref{fig:SNRacc}, we repeated the grid of simulations for the \citet{hirano:2011} RM model instead, with the same assumptions of no limb darkening and zero micro/macro turbulence. Even so, a substantial systematic shift is observed of 20\%. Adding in other complexities, such as limb darkening, turbulence velocities, etc adds further potential offsets.

For these reasons, we do not consider our RM SNR formula to be of the same caliber of the earlier two exact calculations. We did attempt to use the \citet{hirano:2011} grid to derive a separate SNR formulae, but poorer accuracy as well as an extremely convoluted that would have limited practical value. We thus recommend treating our formula as an order-of-magnitude level calculation. Despite this limitation, this appears to be the first expression of its kind and it reveals in a compact analytic form the broad dependencies of the input parameters on detectability. We suggest those looking to debias a population of RM detections will need to use full numerical SNR calculations, but should expect similar trends in detection bias seen in our expression.
	
\section*{Acknowledgments}
We thank Songhu Wang for helpful discussions over the course of this
work.
D.K. thanks donors Mark Sloan,
Douglas Daughaday,
Andrew Jones,
Elena West,
Tristan Zajonc,
Chuck Wolfred,
Lasse Skov,
Graeme Benson,
Alex de Vaal,
Mark Elliott,
Stephen Lee,
Zachary Danielson,
Chad Souter,
Marcus Gillette,
Tina Jeffcoat,
Jason Rockett,
Tom Donkin,
Andrew Schoen,
Jacob Black,
Reza Ramezankhani,
Steven Marks,
Nicholas Gebben,
Mike Hedlund,
Dhruv Bansal,
Jonathan Sturm,
Rand Corp.,
Leigh Deacon,
Ryan Provost,
Brynjolfur Sigurjonsson,
Benjamin Paul Walford,
Nicholas De Haan,
Joseph Gillmer,
Emerson Garland,
Alexander Leishman,
Queen Rd. Fnd. Inc,
Brandon Pearson,
Scott Thayer,
Benjamin Seeley,
Frank Blood,
Ieuan Williams,
Jason Smallbon,
Xinyu Yao \&
Axel Nimmerjahn.
D.K. acknowledges support from NASA Grant \#80NSSC23K1313 and \#80NSSC21K0960.

\section*{Data Availability}

The grid of models used to generate the RM effect are available at \wwwcoolworlds.

\appendix

\bsp
\label{lastpage}
\end{document}